\documentclass[acus]{JAC2003}

\usepackage{graphicx}
\usepackage{booktabs}
\usepackage{placeins}

\usepackage{fixltx2e}

\author{G.~Iadarola$^{(1),(2)}$ and  G.~Rumolo$^{(1)}$\\
$^{(1)}$ CERN, Geneva, Switzerland, $^{(2)}$ Universit\`{a} di Napoli ``Federico II'', Naples, Italy}

\begin{document}
\title{PyECLOUD and build-up simulations at CERN}

\maketitle

\begin{abstract}
  PyECLOUD is a newly developed code for the simulation of the electron cloud (EC) build-up in particle accelerators. Almost entirely written in Python, it is mostly based on the physical models already used in the ECLOUD code but, thanks to the implementation of new optimized algorithms, it exhibits a significantly improved performance in accuracy, speed, reliability and flexibility. Such new features of PyECLOUD have been already broadly exploited to study EC observations in the Large Hadron Collider (LHC) and its injector chain as well as for the extrapolation to high luminosity upgrade scenarios. 
\end{abstract}

\section{INTRODUCTION}

The analysis of the electron cloud (EC) observations in the Large Hadron Collider (LHC) and its injectors have raised new challenges for the EC build-up simulations. On one hand, for a correct understanding of machine observations it is often necessary to deal with beams with thousands of bunches and with non idealities like non uniform bunch populations and bunch lengths along the train. On the other hand, the demand for extensive parametric scans gives quite stringent requirements in terms of speed and reliability.  

CERN has a long experience in the EC build-up simulation, mostly carried out with the ECLOUD code, developed and maintained at CERN since 1997 \cite{ecl_zimm1, ecl_zimm2, ECLDcode, RRZ}. Unfortunately, due to its not modular structure and to the programming language (FORTRAN 77), this code would need a deep reorganization and a serious upgrade to enable it to fulfill the aforementioned requirements.

Therefore we have decided to write a fully reorganized code, in a more modern and powerful language, considering that the initial effort would be compensated by a significantly increased efficiency in future developments and debugging. The new code has been called PyECLOUD, since it is almost entirely written in Python and is largely based on the physical models of the ECLOUD code. Nevertheless, several features and implementations have been modified, in some cases completely redesigned, with respect to ECLOUD, with substantial improvements in terms of reliability, accuracy, speed and usage flexibility.

\section{PyECLOUD}
Like ECLOUD, PyECLOUD is a 2D macroparticle (MP) code, where the electrons are grouped in MPs in order to achieve a reasonable computational burden.

\begin{figure}[!h]
   \centering
   \vspace*{-2mm}
   \includegraphics[width=75mm]{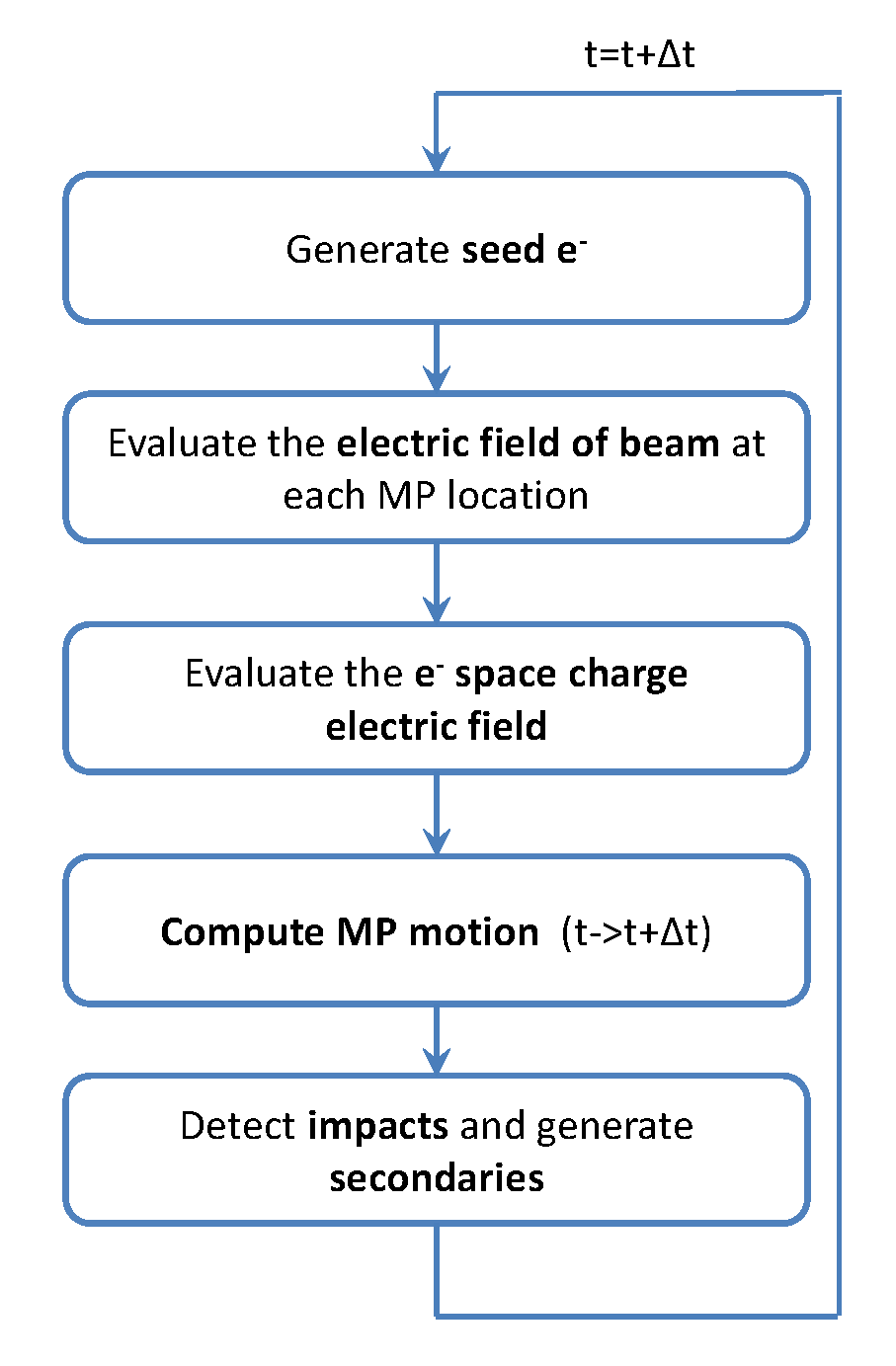}
   \vspace*{-2mm}
   \caption{Flowchart representing PyECLOUD main loop.}
   \vspace*{-2mm}
   \label{fig:flowch}
   
\end{figure}

The dynamics of the MP system is simulated following the flow diagram sketched in Fig.~\ref{fig:flowch}. \\
\begin{figure*}[htb]
    \centering
    \includegraphics*[width=168mm]{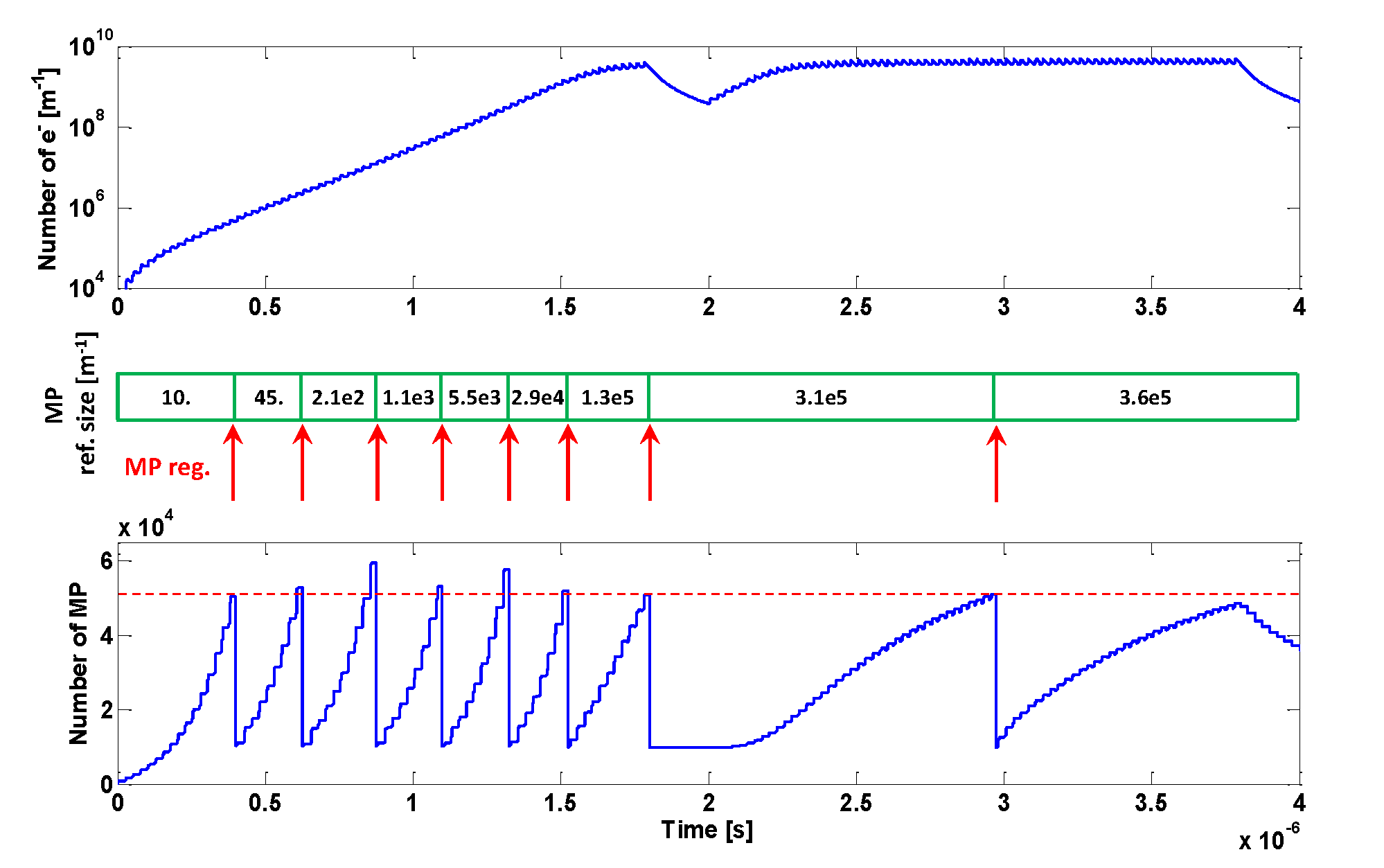}
    \caption{Top: evolution of the number of electrons in the beam pipe for an LHC type beam with 25ns bunch spacing in the SPS (2 trains of 72 bunches); middle: evolution of the reference MP size; bottom: evolution of the number of MPs, the regeneration threshold is highlighted in red.}
    \label{fig:regen}
\end{figure*}
At each time step,  
\textit{seed} electrons, due to residual gas ionization and/or to synchrotron radiation induced photoemission from the chamber walls, are generated with a number consistent with the passing beam slice and with positions and momenta determined by theoretical or empirical models.\\
Then the electric field acting on each MP is evaluated: the field of the beam is precomputed on a suitable rectangular grid, loaded from file and obtained at each MP location by a linear (4 points) interpolation; the \textit{space charge} contribution of the electron system itself is calculated by a classical Particle in Cell (PIC) algorithm, where the finite difference method is employed to solve the Poisson equation with perfectly conducting boundary conditions on the beam chamber.\\
Once the total electric field at each MP location is known, MP positions and momenta are updated by integrating the dynamics equation; at this stage the presence of an externally applied dipolar magnetic field can also be taken into account.\\
At each time step, a certain number of MPs can hit the wall. In these cases a proper model of the secondary emission process is applied to generate charge, energy and angle of the emitted electrons. According to the size of the emitted charge, a rescaling of the impinging MP can be performed or new MPs can be emitted.

\subsection{MP size management}

One of the peculiarities of the EC build-up process is the fact that, due to the multipacting effect, the number of electrons can spread several orders of magnitude during the passage of the bunch train (see Fig.~\ref{fig:regen} - top). As a consequence, it is impossible to choose a MP size which is suitable for the entire simulation, allowing both a satisfactory description of the phenomena and a computationally affordable number of MPs. The MP size management in PyECLOUD has been significantly improved with respect to ECLOUD and will be briefly described in this subsection.

MP sizes are not enforced throughout the simulation process but are determined step by step by ``decisions" taken during the execution. For this purpose a target MP size $N_{ref}$, dynamically adapted during the simulation, is employed to control the number of electrons per MP. In particular:
\begin{itemize}
	\item The size of MPs generated by \textit{seed} mechanisms is exactly $N_{ref}$;
	\item When a MP hits the wall, it is simply rescaled according to the Secondary Electron Yield (SEY) if the emitted charge is below $1.5N_{ref}$, otherwise "true" secondary MPs are generated so that the resulting MP size is as close as possible to $N_{ref}$;
	\item Once per bunch passage, a cleaning procedure is performed, which deletes the MPs with charge lower than $10^{-4}N_{ref}$.
\end{itemize}

$N_{ref}$ is changed whenever the total number of MPs becomes larger than a certain threshold defined in the input file(typical value $\sim 10^5$), which means that the computational burden has become too high. When this happens, a \textit{regeneration} of the set of MPs is applied, by the following procedure (see Fig.~\ref{fig:regen}):

\begin{itemize}
\item Each MP is assigned to a cell of a uniform grid in the 5-D phase space $(x,y,v_x,v_y,v_z)$ obtaining the phase space distribution of the electron gas distributed on the mesh points;
\item The new $N_{ref}$ is chosen in order to get a target number of MPs (typically 5-10 times smaller than the regeneration threshold), which still allows for an accurate simulation but with a more reasonable computational effort;
\item A new set of MPs, having the new reference size, is generated according to the computed distribution.
\end{itemize}

The preservation of the entire phase space is very important in  EC build-up simulation since the dynamics imparted by passing bunches generates very distinctive velocity distributions at the different time steps and the conservation of few specific moments would not guarantee a sufficient accuracy.\\
Several numerical test have shown that the errors on the total charge and the total energy which are introduced by this procedure, are about 1\% at the first time step after the regeneration and they become even smaller at the first bunch passage after the regeneration. 


\begin{figure}[!h]
   \centering
   \vspace*{-2mm}
   \includegraphics[width=85mm]{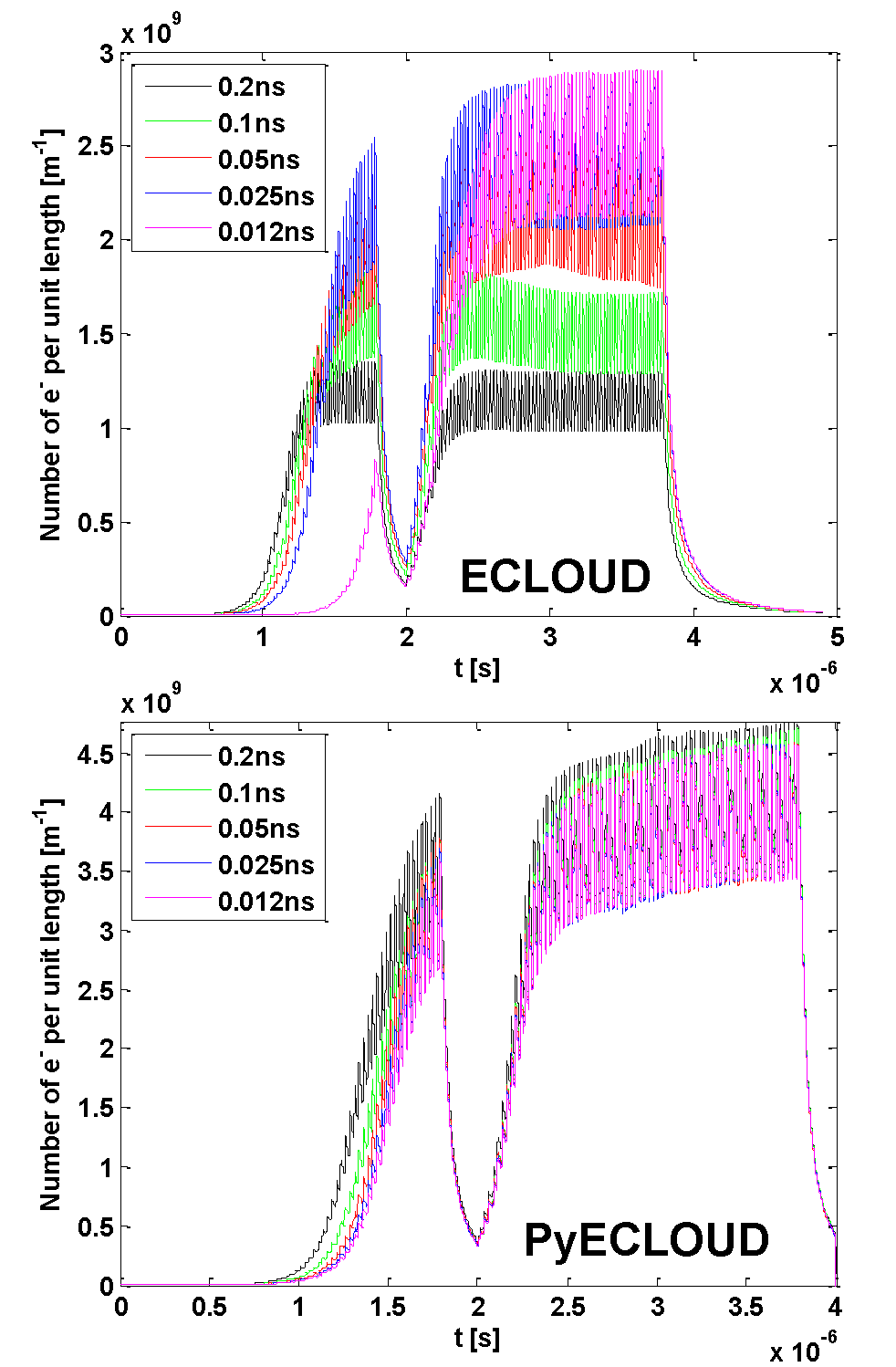}
   \vspace*{-2mm}
   \caption{Electron cloud build-up simulations for different time steps in ECLOUD (top) and PyECLOUD (bottom). Simulated case: SPS MBB bending magnet, 26GeV, two trains of 72 bunches with 225ns gap, 25ns bunch spacing, 1.1$\cdot10^{11}$ protons per bunch.}
   \vspace*{-2mm}
   \label{fig:pyecloud_convergence}
\end{figure}

\begin{table}[hr]
  \centering
  \small
\begin{tabular}{lrr}
\toprule
 & \multicolumn{2}{c}{Processing time}\\
Time step [ps]&ECLOUD& PyECLOUD  \\
\midrule
200 & 29 min & 12 min\\
100 & 1 h 27 min & 13 min\\
50 & 1 h 45 min & 24 min\\
25 & 3 h 7 min & 40 min\\
12 & 4 h 15 min & 1 h 6 min\\ 
\bottomrule
\end{tabular}
\caption{Computation time required by ECLOUD and PyECLOUD for the simualtions in Fig.~\ref{fig:pyecloud_convergence}.}
  \label{tab:Perf}
\end{table}

\subsection{Performances}

The passage from ECLOUD to PyECLOUD had a significant impact on the performances both in terms of accuracy and of computational efficiency.
Fig.~\ref{fig:pyecloud_convergence} shows a comparison between the two codes in terms of convergence properties with respect to the time step which is chosen for the simulation. While in ECLOUD it is quite difficult to achieve a good convergence, PyECLOUD gives a good estimate of the total number of electrons in the chamber already for quite large time steps ($\sim 0.1$ns) while a satisfactory convergence is obtained for a time step of the order of $25$ps.

For the same test cases the simulation time required by the two codes is reported in Tab.~\ref{tab:Perf} showing that the improvements introduced in PyECLOUD had also a positive impact on the code efficiency.

Furthermore, the new code has been designed in order to offer an increased usage flexibility, allowing to deal with irregular beam structures e.g. non uniform bunch intensity and/or bunch length along the bunch train, irregular bunch spacings and bunch profiles.

Thanks to these new features, PyECLOUD has been already largely exploited at CERN for several EC simulation studies for the LHC and its injector chain \cite{ECLOUD12_gr, ipac12_hm, ipac12_fritz, ECLOUD12_cb}. In particular, as described in detail in \cite{ICAP12_gi}, PyECLOUD simulations have been used to reconstruct the evolution of the SEY of the chambers in the LHC arcs, from the measurement of the heat load deposited on the beam screen of the cryogenic magnets. The new code also allows us to estimate the bunch by bunch energy loss due to the interaction of the beam with the EC and to export the electron distribution seen by each bunch. The first feature allowed us to benchmark the results against bunch by bunch stable phase measurements \cite{ECLOUD12_jem} while the second was used, together with HEADTAIL simulations, to analyze the instabilities observed in the LHC with 25ns bunch spacing \cite{ECLOUD12_hb}.  

Simulation studies have also addressed the EC formation in the common vacuum chambers of the LHC. An example of this kind of application will be described in the following section in order to show the capability of PyECLOUD to deal with beams made of thousands of bunches with irregular spacings.

\section{EC build up in LHC common vacuum chambers}

\begin{figure}[!h]
   \centering
   \vspace*{-2mm}
   \includegraphics[width=90mm]{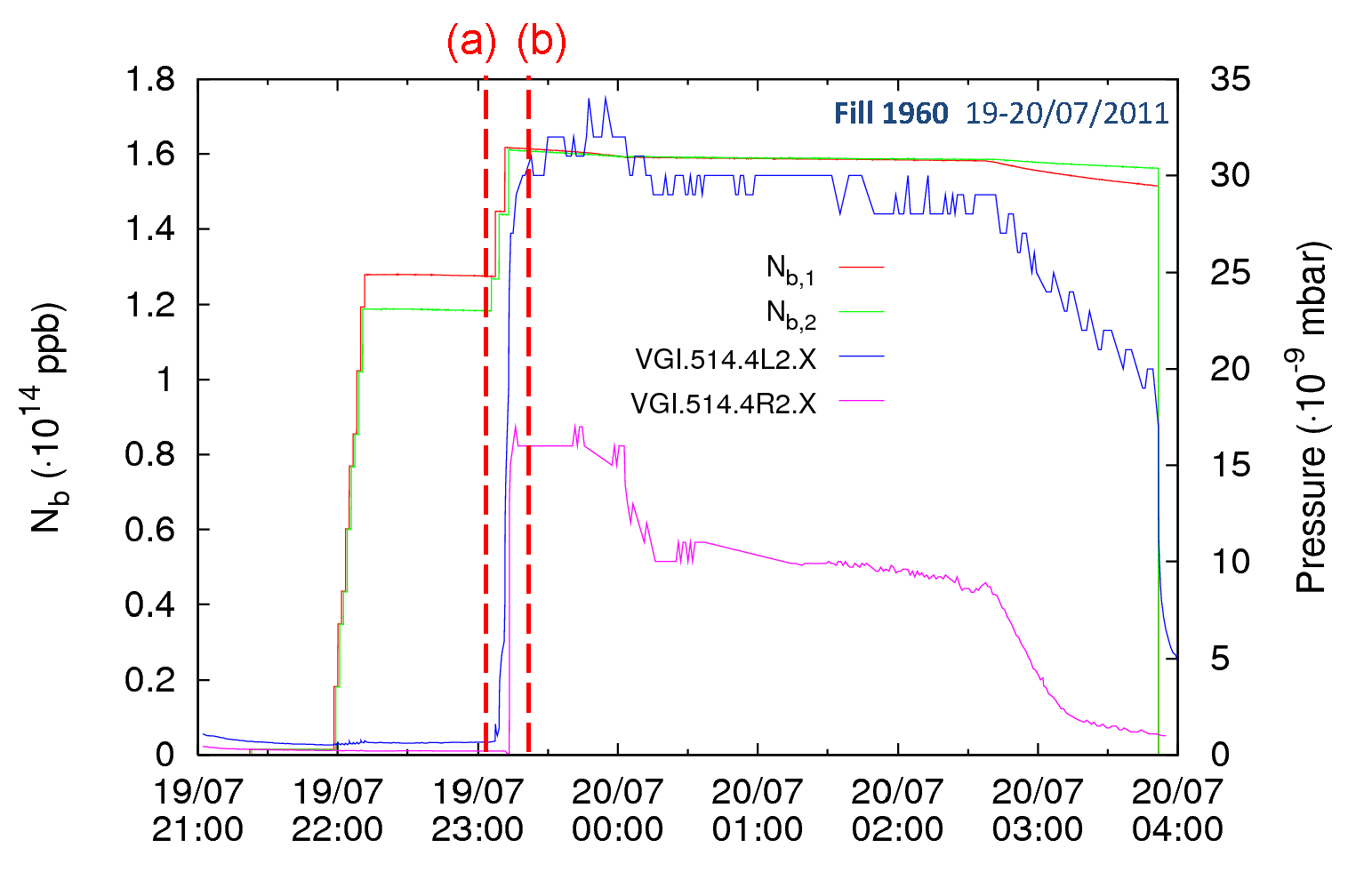}
   \vspace*{-2mm}
   \caption{Pressure evolution in the 800mm chambers near the ALICE experiment of the LHC during a proton physics fill. The total beam intensity is the two rings is also included.}
   \vspace*{-2mm}
   \label{fig:alice_press}
\end{figure}

Common vacuum chambers having 800mm diameter are installed on both sides of the ALICE experiments in the Long Straight Section 2 (LSS2) of the LHC. During 2011 operation with 50ns bunch spacing an important pressure rise was noticed in these chambers with a significant impact on the background observed by the ALICE experiment.\\
The analysis of the pressure data has shown that a severe pressure increase is observed only when the two rings of the LHC are completely filled. Fig.~\ref{fig:alice_press} shows the pressure evolution during a proton-proton physics fill (with 50ns spacing) in which the injection from the SPS of the last two trains of 144 bunches was delayed by about one hour. It can be noticed that the pressure rise appears already at the injection energy (450GeV), but only after the last two injections have taken place.

\begin{figure}[!h]
   \centering
   \vspace*{-2mm}
   \includegraphics[width=85mm]{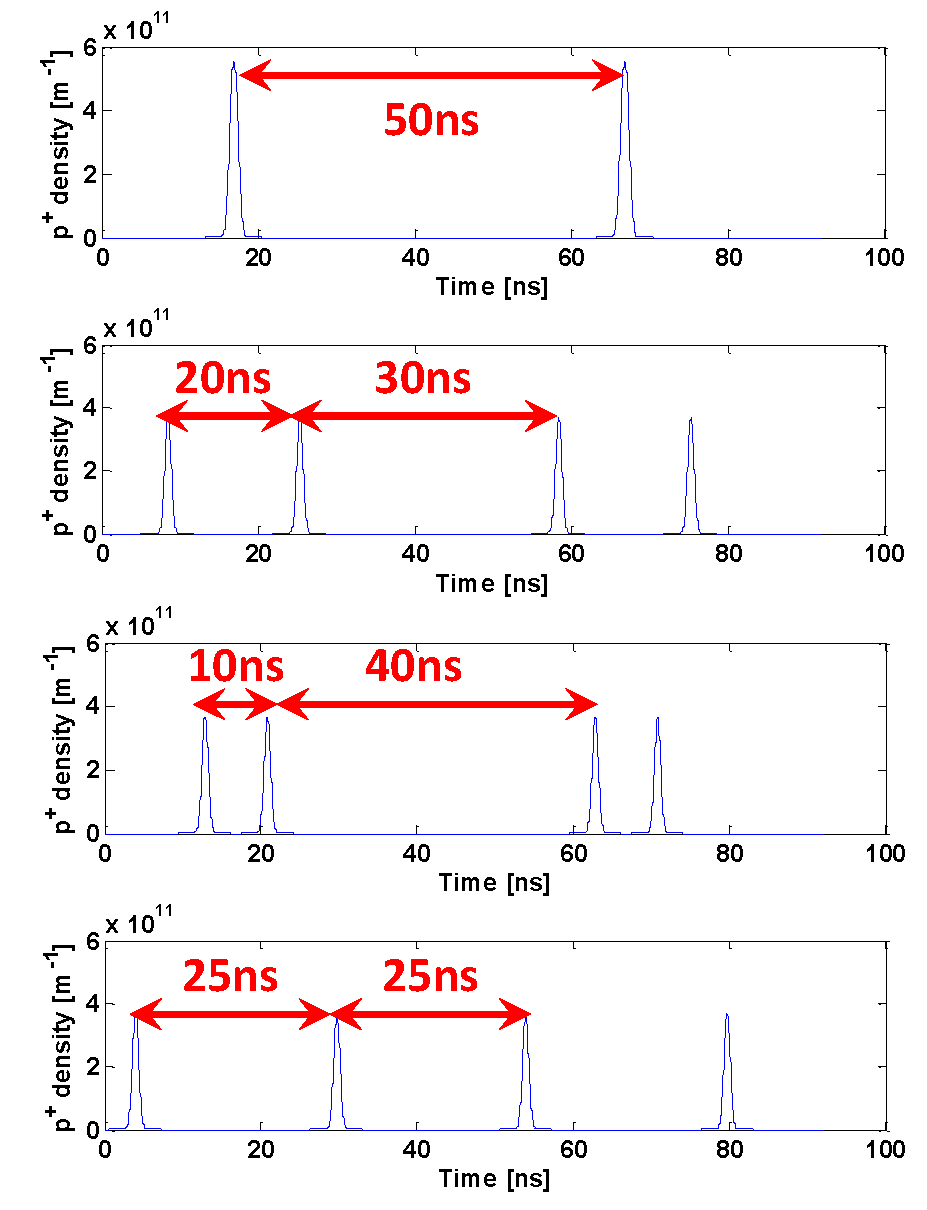}
   \vspace*{-2mm}
   \caption{\textit{Hybrid} bunch spacings which are observed at different sections of the 800mm common chambers.}
   \vspace*{-2mm}
   \label{fig:b_spac_alice}
\end{figure}

\begin{figure}[!h]
   \centering
   \vspace*{-2mm}
   \includegraphics[width=85mm]{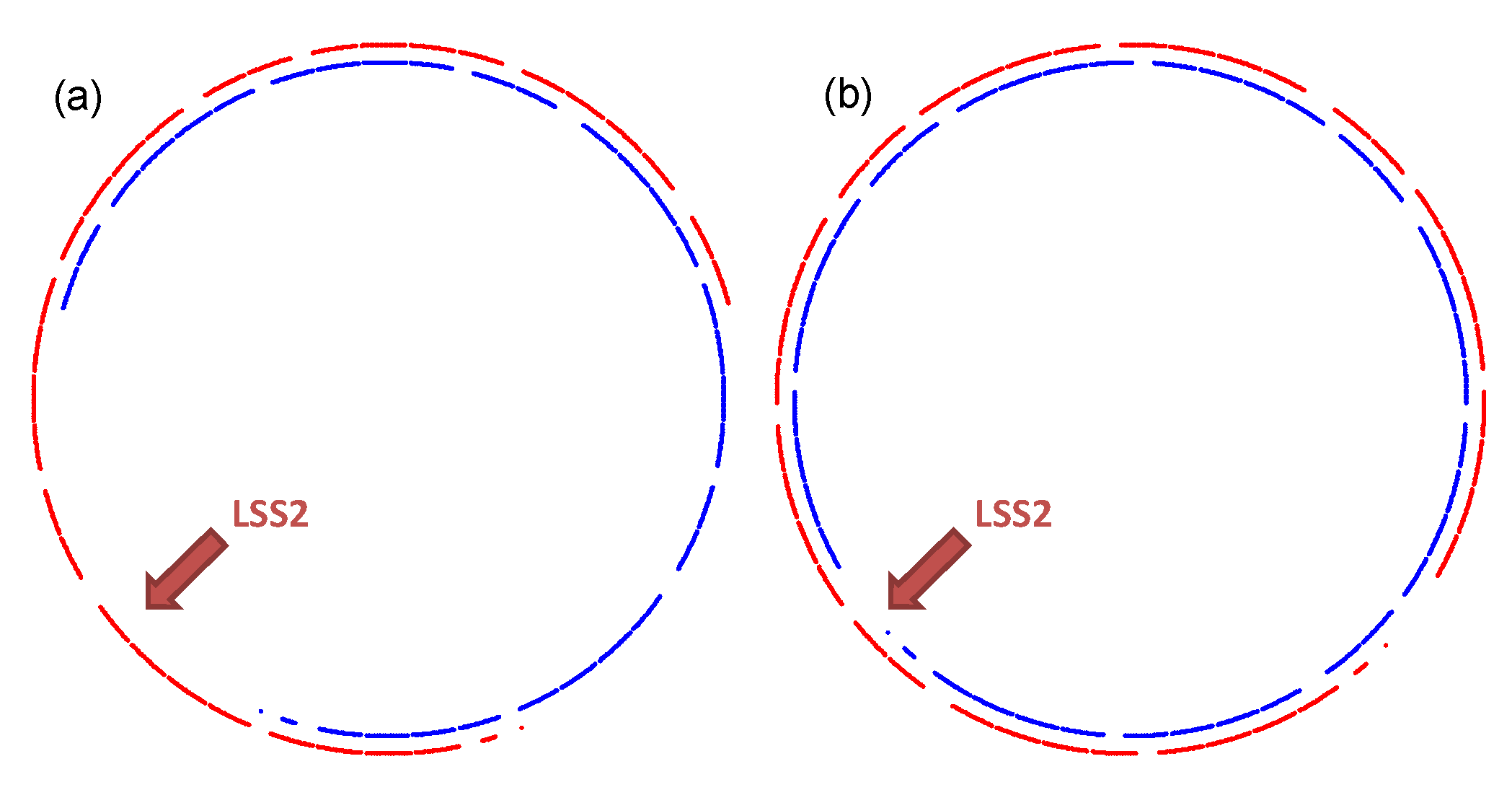}
   \vspace*{-2mm}
   \caption{Filling pattern of the two beams in the LHC before (a) and after (b) the injections of the last two trains. Beam 1 (blue) is clockwise rotating, beam 2 (red) counterclockwise. The position the straight section 2 (ALICE) is highlighted}
   \vspace*{-2mm}
   \label{fig:alice_rings}
\end{figure}

\begin{figure*}[htb]
    \centering
    \includegraphics*[width=168mm]{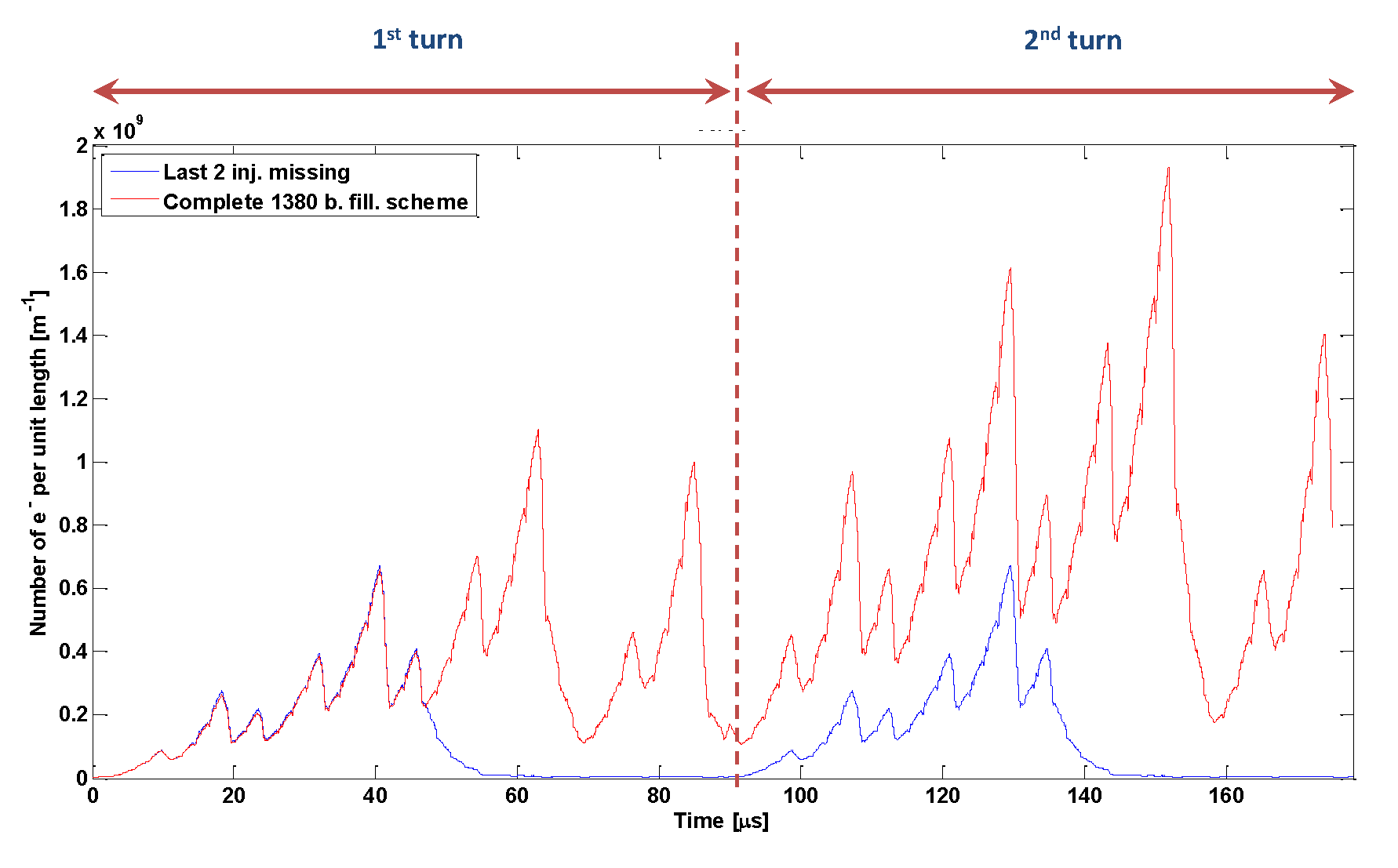}
    \vspace{-2mm}
    \caption{PyECLOUD simulation of the electron cloud build-up in the 800mm common chambers for the filling patterns in Fig.~\ref{fig:alice_rings}.}
    \label{fig:alice_simulation}
\end{figure*}

PyECLOUD simulations have been run in order to investigate if the EC formation in these chambers could explain this pecular behavior. The presence of both counter-rotating beams in the chamber had to be taken into account since it determines different ``hybrid" bunch spacings at the different sections of the $\sim$30m long vacuum chambers when both beams are circulating in the LHC (see Fig.~\ref{fig:b_spac_alice}).\\
In particular the two beam configurations in Fig.~\ref{fig:alice_rings} have been simulated, which correspond to the beam patterns in the two rings at the moments indicated by (a) and (b) in Fig.~\ref{fig:alice_press}. The results at a certain section of the considered vacuum pipes are shown in Fig.~\ref{fig:alice_simulation}.\\ 
It can be noticed that, in the configuration of Fig.~\ref{fig:alice_rings}a, both beams present a gap of about one quarter of the length of the ring. Probably due to the quite large radius, the EC can develop only when both beams are passing in the chamber, while a decay of the number of electrons is  observed when only one beam is passing and the gap is long enough to allow a complete reset of the EC between subsequent turns. \\
On the other hand, after the injection of the last two trains from the SPS, the layout of the beam in the two rings looks like the one in Fig.~\ref{fig:alice_rings}b where no large gap is present in any of the two beams. As a consequence a complete decay of the EC between subsequent turns is not possible anymore. In fact, a memory effect is observed between turns with a strong enhancement of the EC activity and, as consequence, of the electron stimulated gas desorption leading to the observed pressure rise. 

\section{CONCLUSIONS}

PyECLOUD is a new EC build-up simulation code, which has been developed at CERN for a reliable and efficient analysis of the EC observations in LHC as well as for EC studies related to the high luminosity upgrade of the LHC and its injectors.

Based on the physical models already implemented in ECLOUD, the new code features several improvements in terms of algorithms and implementation (e.g. a new MP size management method) with a significant impact on speed and convergence properties. Moreover the code is explicitly designed to deal with irregular beam structures (e.g. different bunch intensities and bunch lengths along the train, arbitrary spacings and profiles) in order to allow for an accurate analysis of EC observations in CERN accelerators. 

Several EC studies for the LHC and its injectors have been already carried out with the new code giving very encouraging indications on the reliability of the models and numerical solutions.

\section{ACKNOWLEDGMENTS}
The authors would like to express their gratefulness to  G. Arduini, H. Bartosik, C. Bhat, V. Baglin, R. De Maria, O. Dominguez, M. Driss Mensi, J. Esteban-Muller, K. Li, H. Maury Cuna, G. Miano, E.~Mé\'{e}tral, H. Neupert, 
G. Papotti, E. Shaposhnikova, M. Taborelli, L. Tavian, C. Y. Vallgren, and F. Zimmermann for the support they 
provided in the code development, the simulation work and the machine data analysis as well as for valuable comments and discussions.

\end{document}